\documentclass{elsart}
\journal{Ocean Modelling}
\usepackage{mytex}

\begin{document}
\bibliographystyle{named}

\begin{frontmatter}
\title{
Identification of Optimal  Topography by Variational Data Assimilation. }

\author{ Eugene Kazantsev}
\address{ INRIA, projet MOISE, \\
 Laboratoire Jean Kuntzmann,\\
BP 53\\
38041 Grenoble Cedex 9 \\
France}
\ead{kazan@imag.fr}

\begin{abstract} 
The use of the data assimilation technique to identify optimal topography is discussed  in frames of time-dependent motion governed by non-linear  barotropic ocean model. 
Assimilation of artificially generated data allows to measure the influence of
 various error sources and to classify the impact of noise that is present in observational data and model parameters. The choice of assimilation window is discussed.
 Assimilating noisy data with longer windows provides higher accuracy of identified topography. The topography identified once by data assimilation can be successfully  used for other model runs that start from other initial conditions and are situated in other parts of the model's attractor.
\end{abstract}
\begin{keyword} 
Variational Data Assimilation; Bottom Topography; Barotropic ocean model;
\end{keyword}

\end{frontmatter}

\maketitle

\picstoend

\section{Introduction.}

It is now well known, even the best model is not sufficient to make a forecast. Any model depends on a number of parameters, it requires initial and boundary condition and other data that must be collected and used in the model.  However, interpolating data from observational points 
 to the model grid or smoothing of observed data is not an optimal way to incorporate these data in a model. Lorenz, in his  pioneering work  \cite{Lor63} has shown that a geophysical fluid is extremely sensitive to initial conditions. This fact requires to bring  the model and its initial data together, in order to work with the couple "model-data" and identify the optimal initial data for the model taking into account simultaneously the information contained in the observational data and in the equations of the model. 

Optimal control methods \cite{Lions68} and perturbations theory   \cite{Marchuk75} applied to the data assimilation technique (\cite{Ledimet86}, \cite{ldt86}) show  ways to do it. They allow to retrieve an optimal initial point for a given model from heterogeneous observation fields. Since early 1990's many mathematical and geophysical teams have been involved in the development of the data assimilation strategies. One can cite many papers devoted to this problem, as  in the domain of development of different techniques for the  data assimilation such as nudging (\cite{Verron92}, \cite{VH89}, \cite{Blayo94}, \cite{AB08})  ensemble me\-thods and Kalman filtering (\cite{bennett87}, \cite{Cohn97}, \cite{Pham98}, \cite{Bras03}), variational data assimilation (3DVAR and 4DVAR) (\cite{ldt86}, \cite{ld85}, \cite{thepautcourtier91}, \cite{navon92}) and in the domain of its applications to the atmosphere and oceans.  

These methods have proved capable of combining information from the model  and the heterogeneous set of available observations. They have supported a remarkable increase in forecast accuracy (see, for example  \cite{Kalnay90})
The success of the data assimilation  stipulates the development of modern models together with  methods allowing to integrate  all  available data in the model. Thus, in 1997 acknowledging the need for better ocean observations and ocean forecasts and with the scientific and technical opportunity that readily available satellite data had delivered, the Global Ocean Data Assimilation Experiment (GODAE) was initiated to lead the way in establishing global operational oceanography. Another example, the  Mercator Ocean  Group  was founded in  2002  to set up an operational system for describing the state of the ocean, an integral part of our environment. Input for the Mercator system comes from ocean observations measured by satellites or in situ observations through measurements taken at sea. These measurements are assimilated by the analysis and forecasting model. The assimilation of observation data in a model is used to describe and forecast the state of the ocean for up to 14 days ahead of time.

However,  majority of the data assimilation methods are now intended to identify and reconstruct an optimal initial point for the model. Since Lorenz \cite{Lor63}, who has pointed out on the importance of precise knowledge of the starting point of the model, essentially  the  starting point is considered as the control parameter and the target of the data assimilation. 
 
Of course, the model's flow  is extremely sensitive to its initial point. But, it is reasonable to suppose that an ocean  model is also sensitive to many other parameters, like bottom topography, boundary conditions, forcing fields and friction coefficients. All  these parameters and values  are also extracted in some way  from observational data, interpolated on the model's grid and can neither be considered as exact, nor as optimal for the model. On the other hand, due to nonlinearity and intrinsic instability of model's trajectory, its sensitivity to all these external parameters may also be exponential. 

Numerous studies show strong dependence of the model's flow on the boundary data (\cite{VerronBlayo}, \cite{Adcroft98}), on the representation of the bottom topography (\cite{Holland73}, \cite{EbyHolloway},\cite{LoschHeimbach}), on the wind stress (\cite{Bryan95},\cite{Milliff98}), to diffusivity coefficients \cite{Bryan87} and to fundamental parametrisations like Boussinesq and hydrostatic hypotheses \cite{LoschAdcroft}. 

Despite the bottom topography and the boundary configuration of the ocean are steady and can be measured with much better accuracy than the model's initial state, it is not obvious how to represent them  on the model's grid because of the limited resolution. It is known for 30 years, that requiring the large scale ocean flow to be well represented, one have to smooth the topography to get only corresponding large-scale components of relief  \cite{Ilin}. In this case, the influence of subgrid-scales has to be parameterized. But it is not clear how to apply the parametrisation for a given model with a given resolution. 
It is shown in  \cite{Penduff02}, that different smoothing of the topography pattern may significantly  change the model's properties.

This paper is devoted to  the use of the data assimilation procedure in order to identify an optimal bottom topography in a simple barotropic ocean model.  The first attempt to use the topography as a control parameter was performed in \cite{LoschWunsch} for a steady solution of a linear shallow-water model in a zonal channel. The control parameter in the data assimilation procedure is the parameter that is  modified  to bring the model within an estimated error of the observations. It was shown that the relationship between topography and the surface elevation does not have an unique inverse and, hence, all the details of the depth field can not be recovered. In this paper we work also with a simple, but different model. First, the model's solution  is not stationary. We control the topography in the time dependent motion. And second, the model is nonlinear with chaotic intrinsicly unstable behavior. Thus, our study is placed in the usual context of  data assimilation when the dependence of the model's flow on the controlled parameter is strong. In addition, barotropic ocean model is a well studied toy model frequently used for a preliminary study. 

 The sensitivity of this model to perturbations of the bottom topography has been described in \cite{toposens}. It was shown the model's sensitivity to topography differs from the sensitivity to initial state at short time scales only. But, when considering scales longer than 3 days, the sensitivity of the solution becomes the same to any source of perturbation. Intrinsic model's instability dominates at these time scales and the source of the perturbation is no longer important. No matter how the perturbation is introduced in the model, in several days the growth  becomes exponential in time with the  rate determined by the model's dynamics independently on the source of the perturbation.
 
 Another important feature for data assimilation consists in the existence of  a kernel of the sensitivity operator. That means there is no way to reconstruct the exact topography pattern by assimilation because of the presence of modes the flow is not sensitive at all. This fact must be taken into  account in the data assimilation analysis.

The paper is organized as follows. The second section describes the model, its adjoint and the data assimilation procedure. The third section is devoted to numerical experiments and discussion.

\section{The Model}

We consider the shallow-water model with the rigid lid assumption
\beqr
\der{u}{t} - fv + u\der{u}{x}+v\der{u}{y}  +\fr{1}{\rho_0} \der{p}{x} &=& \frac{\tau^{(x)}}{\rho_0 H_0} -\sigma u + \nu \Delta u\nonumber\\
\der{v}{t} + fu +  u\der{v}{x}+v\der{v}{y} +\fr{1}{\rho_0}\der{p}{y} &=& \frac{\tau^{(y)}}{\rho_0 H_0} -\sigma v + \nu \Delta v \label{sw}\\
\der{(H u)}{x} + \der{(H v)}{y} &=& 0 \nonumber
\eeqr
where $\rho_0$ is the mean density of water and $H_0$ is the characteristic depth of the basin. The Coriolis parameter $f$ is supposed to be linear in $y$ coordinate: $f=f_0+\beta y$.

The third equation allows us to introduce the streamfunction $\psi$, such as 
\beq
 Hu=-\der{\psi}{y},\hspace{5mm} Hv=\der{\psi}{x} \label{strf-uv}
\eeq
Denoting the vorticity by $\omega=\der{v}{x}-\der{u}{y}$ we get 
\beq
\omega =  \der{}{x} \fr{1}{H} \der{\psi}{x} + \der{}{y} \fr{1}{H}\der{\psi}{y} \label{stat}
\eeq 
This equation possesses a solution when $H(x,y)$ is always positive. Numerically, it can be solved by Cholesky decomposition method, for example. 

Using this notation we calculate the curl of the first two equations of the system~\rf{sw}. We get 
\beq
\der{\omega}{t} +(\omega+f) div \vec{u} +u\der{(\omega+f)}{x}+v\der{(\omega+f)}{y} = \nu\Delta\omega -\sigma\omega +  \frac{{\cal F}(x,y) }{\rho_0 H_0} \nonumber 
\eeq
or 
\beq
\der{\omega}{t} +\jac(\psi, \fr{\omega+f}{H} )= \nu\Delta\omega -\sigma\omega + \frac{{\cal F}(x,y) }{\rho_0 H_0} \label{btp} 
\eeq
where $\jac(\psi,\omega)=\der{\psi}{x}\der{\omega}{y}-\der{\psi}{y}\der{\omega}{x}$ is the Jacobian operator and ${\cal F}(x,y)=-\fr{\partial \tau_x}{\partial y} +
 \fr{\partial \tau_y}{\partial x}$. 

The system \rf{btp} is considered in a bounded domain $\Omega$ and is subjected to the impermeability and slip boundary conditions:
\beq 
\psi\mid_{\partial\Omega} =0,\;\omega\mid_{\partial\Omega} =0 \label{bcref}
\eeq

The model is discretised in space by finite elements method. Details of the discretisation can be found in \cite{toposens}.

This system has been forwarded in time by the following scheme, 
\beqr
 \fr{ \omega^{n+1}-\omega^{n-1}}{2\tau} &+& 
 \jac( \psi^{n}, \fr{\omega^{n}+ f_0+\beta y}{H}) 
 =-\mu \Delta \fr{ \omega^{n+1}+\omega^{n-1}}{2} - \label{sch} \\
&-&\sigma \fr{ \omega^{n+1}+\omega^{n-1}}{2} 
+  \frac{{\cal F}(x,y) }{\rho_0 H_0} \nonumber
\eeqr
The first step is performed using the two stage process. On the first stage we calculate the value of $\omega^{1/2}$ at the time $\tau/2$. At the second stage we use this value to calculate  $\delta\omega^{1}$ with the accuracy  of second order at the time $\tau$.
\beqr
 \fr{ \omega^{1/2}-\omega^{0}}{\tau} + 
 \jac( \psi^{0}, \fr{\omega^{0}+ f_0+\beta y}{H}) 
 =-\mu \Delta \fr{ \omega^{1/2}+\omega^{0}}{2}
-\sigma \fr{ \omega^{1/2}+\omega^{0}}{2} 
+  \frac{{\cal F}(x,y) }{\rho_0 H_0} \nonumber \\
 \fr{ \omega^{1}-\omega^{0}}{\tau} + 
 \jac( \psi^{1/2}, \fr{\omega^{1/2}+ f_0+\beta y}{H}) 
 =-\mu \Delta \fr{ \omega^{1}+\omega^{0}}{2}
-\sigma \fr{ \omega^{1}+\omega^{0}}{2} 
+  \frac{{\cal F}(x,y) }{\rho_0 H_0} \label{1step}
\eeqr
This procedure helps us to avoid numerical oscillations at the beginning of the integration of the  model.

\subsection{Tangent linear and adjoint models}

Let us suppose the couple $\psi(x,y,t),\omega(x,y,t)$ is a solution of the system \rf{stat}, \rf{btp} with a given topography $H= H(x,y)$. If we perturb the topography by some small $\delta H$, we get another solution of the system  $\{\psi+\delta\psi, \omega+\delta\omega\}$. 

Our purpose is to define the relationship between $\delta H$ and $\delta\omega$ supposing both of them to be sufficiently small: 
$$\norme{\delta H} \ll \norme{H} \mbox{ and } \norme{\delta \omega} \ll \norme{\omega}$$ 

We start from the stationary equation \rf{stat}. So far, the  couple $\{\psi+\delta\psi, \omega+\delta\omega\}$ is  a solution of the system with the perturbed topography, it must  satisfy the equation \rf{stat}  
\beq
\omega+\delta\omega =  \der{}{x} \fr{1}{H+\delta H} \der{\psi+\delta\psi}{x} + \der{}{y} \fr{1}{H+\delta H}\der{\psi+\delta \psi}{y} \label{statp}
\eeq 
Using the Taylor development and keeping only linear terms in $\delta H, \; \delta\psi,$ and $ \delta\omega$ we get the equation that allows us to compute $\delta\psi$ from $\delta\omega,\; \delta H$ and the reference streamfunction$ \psi$:  
\beq
\nabla \fr{1}{H}\nabla \delta\psi = \delta\omega+ \nabla \fr{\delta H}{H^2}\nabla \psi  \label{pertpsi}
\eeq

To get the equation for vorticity perturbation, we consider the evolution equation \rf{btp}. As well as above, we write the equation for the perturbed topography using the the Taylor development  and neglect high order terms. 

Skipping  the detailed development of the tangent model, we write it in a short matricial form
 \beq
 \der{\delta\omega}{t} = A(\psi,\omega)\delta\omega+ 
  B(\psi,\omega)\fr{\delta H}{H} -D \delta\omega \label{mateq}
 \eeq
 where operators $A$,  $B$ and $D$  are defined as 
\beqr
A(\psi,\omega) \xi &=& 
-J(\psi,\fr{\xi}{H}) +J( \fr{\omega+ f_0+\beta y}{H}, \biggl(\nabla \fr{1}{H}\nabla\biggr)^{-1} \xi )\label{A} \\
B(\psi,\omega) \xi&=& 
J(\psi,
  \fr{\omega+ f_0+\beta y}{H}\xi)
  + J(\fr{\omega+ f_0+\beta y}{H},\biggl(\nabla \fr{1}{H}\nabla\biggr)^{-1} \biggl(  \nabla \fr{\xi}{H} \nabla \psi\biggr)) \label{B} \\
D \xi &=&   -\nu\Delta\xi +\sigma\xi \nonumber
\eeqr
The  system \rf{mateq} starts from the zero initial state $\delta\omega(x,y,0)=0$ because the purpose is confined to the study of the sensitivity of the solution to the topography, rather than to the initial state. We require the perturbed solution $\{\psi+\delta\psi, \omega+\delta\omega\}$ to have the same boundary condition as $\{\psi, \omega\}$ because we do not want to study the model's sensitivity to boundary conditions. Hence,   perturbations $\delta\psi, \delta\omega$ in equations \rf{pertpsi}, \rf{mateq} must satisfy
$ 
\delta\psi\mid_{\partial\Omega} =0,\;\delta\omega\mid_{\partial\Omega} =0 $

Applying the same  time stepping  scheme as for the reference model we get

  \beq
\delta\omega^{N}= \delta\omega(T) = G(\psi,\omega,T)\fr{\delta H}{H} \label{Gmat}
  \eeq
where $G$ is the product of tangent linear operators on each time step of the model. 

To develop the adjoint  model, we calculate  the adjoint of the $G(\psi,\omega,T)$ matrix. 

Several remarks  can be made on the tangent model formulation. 
First of all, we can see the right-hand-side of the  tangent linear model \rf{mateq} is  composed by two  terms $A$ and $B$ (\rf{A}, \rf{B}).  The first one, $A$, is responsible for the evolution of a small perturbation by the model's dynamics, while the second one, $B$,  determines the way how the uncertainty is introduced into the model. The first term is similar for any data assimilation, while the second one is specific to the particular variable under identification. This term is absent when the goal is to identify the initial point because the uncertainty is introduced only once, at the beginning of the model integration. But, when the uncertainty is presented in the bottom topography, or some other internal model parameter, the perturbation is introduced at each time step of  the model. 

Another difference consists in the fact that the model state of the tangent linear and adjoint models include one supplementary variable. In this paper we have to add the bottom topography as the third variable to the model state.   

\subsection{Minimization}

The purpose of this work is to identify the optimal topography of the ocean model as the field $H^*(x,y)$ that realizes the minimum of the cost function $\costfun$
\beq
\costfun(H^*(x,y)) = \min_{H} \costfun(H(x,y)) = \min_H \int_0^T \norme{\omega(x,y,t,H)-\omega_{obs}(x,y,t)}^2 dt \label{costfun}
\eeq 
where the norm is defined as 
\beq
\norme{\omega(x,y)}=\biggl(\int_\Omega \omega(x,y)^2 dx dy\biggr)^{1/2} \label{norme}
\eeq
Here, $\omega_{obs}(x,y,t)$ is supposed to be the model's variable reconstructed from  observations. The minimization procedure  is devoted to find the  topography, that ensures the closest model's solution to the observational variable. 

To minimize the cost function $\costfun$ we need first to find its gradient. We start from finding the variation of the functional. 
\beqr
\delta \costfun &=& \costfun\biggl[H(x,y)+\delta H(x,y)\biggr]-\costfun(H(x,y)) =
\nonumber\\
&=&
\int_0^T \sph{\omega(t,H)-\omega_{obs}(t),\omega(t,H+\delta H)-\omega(t,H)} dt =
\nonumber\\
&=&
\int_0^T \sph{\omega(t,H)-\omega_{obs}(t),G(t)\delta H} dt =
\nonumber\\
&=&
\sph{\int_0^T G^*(t)(\omega(t,H)-\omega_{obs}(t))dt,\delta H} 
\eeqr
where the scalar product $\sph{a,b}$ is associated with the norm \rf{norme}.

Thus, the gradient of the cost function can be obtained as the integral of the  product of the adjoint operator $G^*$ \rf{Gmat} and the difference between the model state and observations. 

\beq
\nabla\costfun = \int_0^T G^*(t)(\omega(t,H)-\omega_{obs}(t))dt \label{grad}
\eeq
The operator $G^*$ represents the product of time steps of the adjoint model. 
In practice, we do not need  to conserve the matrix $G^*(t)$ in memory.  The multiplication of a vector by the adjoint operator is realized by the  adjoint model integration starting from this vector.

The  minimization procedure used here was developed by Jean Charles Gilbert and  Claude Lemarechal, INRIA \cite{lemarechal}.  The procedure uses the limited memory quasi-Newton method.

\subsection{Domains and   grids }

The domain was chosen to   represent the North Atlantic region. We assume that the domain is  comprised  in the rectangle between $78^0W \ldots 3^0W$ in longitude and
$15^0N \ldots 65^0N$ in latitude. The boundary of the basin corresponds to the 1 km depth isobath of the ocean. 

To obtain the  forcing in this experiment we have used the data set ``Monthly Global Ocean Wind Stress Components" prepared and maintained by the Data Support Section, Scientific Computing Division, National Center for Atmospheric Research. These data have been prepared by the routine described in \cite{hellerman}. From this data set we choose the mean January  wind stress components $\tau_x$ and $\tau_y$ over the North Atlantic based on 1870-1976 surface observations. These data are presented on the $2^0\times 2^0$ grid. 

The forcing  in this experiment is calculated from these data as
\beqr
{\cal F}(x,y) = -\fr{\partial \tau_x}{\partial \varphi} +
 \fr{1}{\cos\varphi} \fr{\partial \tau_y}{\partial \lambda},\nonumber\\ 
\varphi=20^0+y\times 50^0/L,  \lambda= -40^0+\fr{x\times 50^0/L}{\cos\varphi}
\eeqr
where $L=5500$km is the characteristic length of the basin. The spatial configuration of wind tensions $\tau_x$ and $\tau_y$  is presented in \rfg{topo}A.

The bottom topography has been  interpolated from the   ETOPO5 5-minute gridded elevation data \cite{etopo5}.  This topography  is shown in \rfg{topo}B.

\figureleft{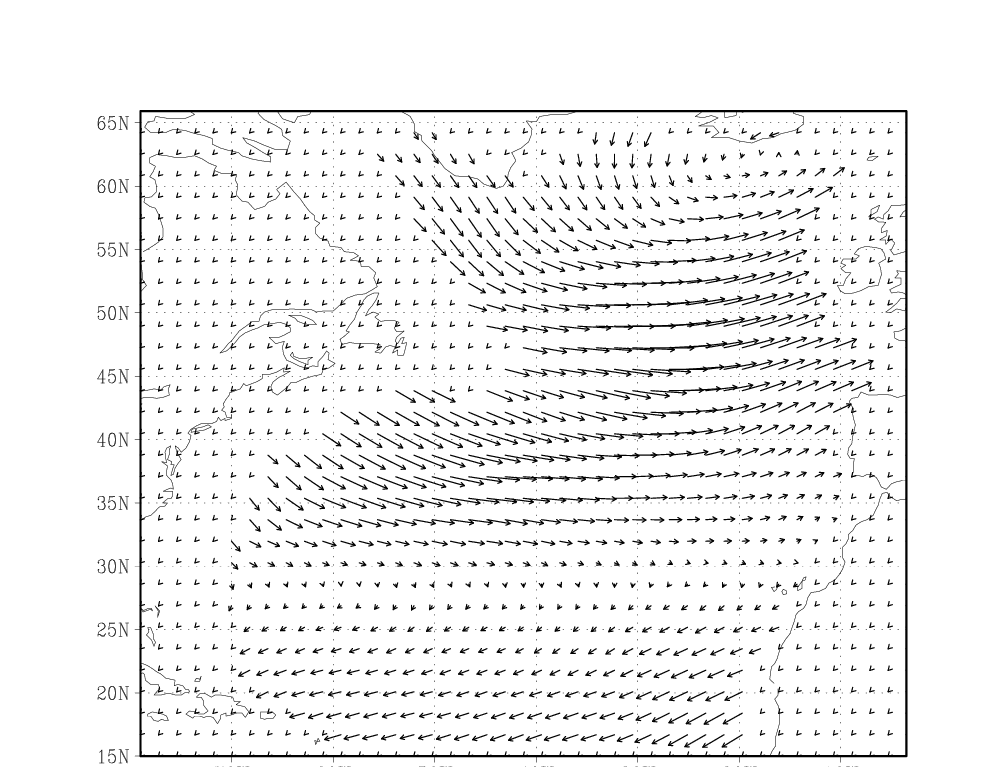}{ Wind tension $\tau_x, \tau_y$ used as the forcing of  the model. The longest arrow   corresponds to $3\fr{dyne}{cm^2}$}{topo}  
\figureright{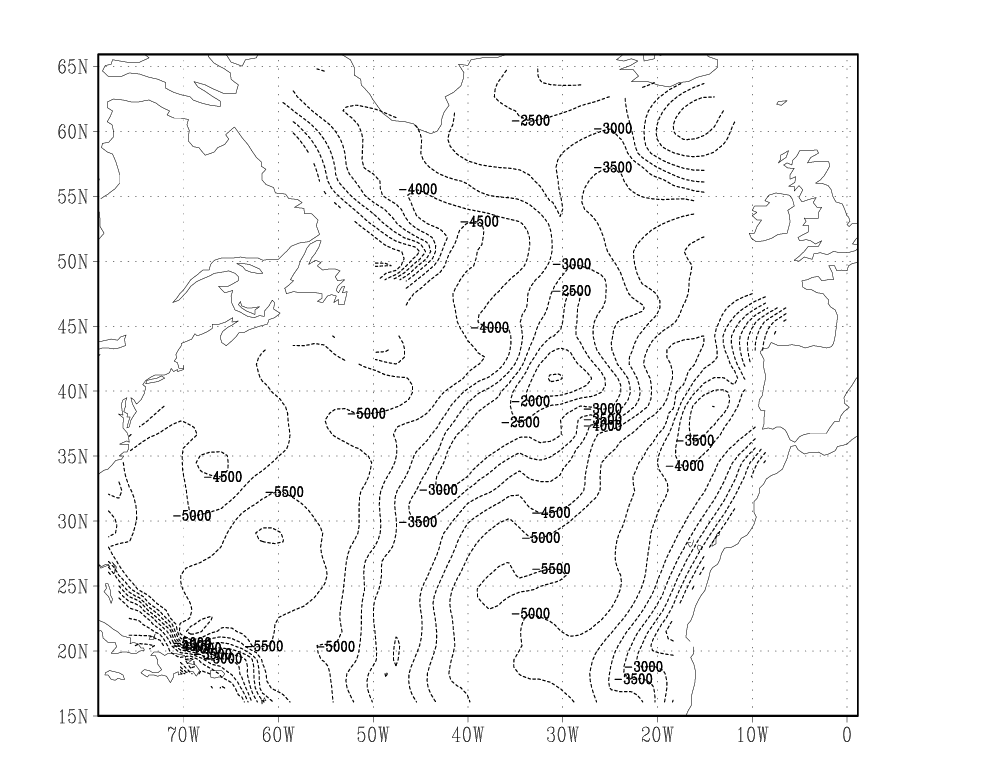}{Model's topography. Contours from -500m to -6500m, interval 500m.} 

The coefficient of Eckman dissipation we choose  as 
$\sigma=5\times10^{-8}s^{-1}.$ The lateral friction coefficient $\mu$ has been chosen in order to avoid numerical instability which occurs due to the concentration of   variability of the model at grid scales. This value has been taken to be $\nu=300\fr{m^2}{s}$, that corresponds to the damping time scale $T_{\nu}= 6$ days for a  wave of 100 km length. 

Forcing and friction coefficients were chosen to ensure chaotic behavior of the model's solution. Data assimilation is especially useful and necessary in situations when  solution strongly depends on the control parameter (initial point, topography or something else).  This strong (exponential in time) dependence has been first pointed out  by \cite{Lor63}. He has shown rapid divergence of trajectories due to model's intrinsic instability which limits the prediction time and leads to chaotic behavior of the model's solution. He has pointed out the real processes in the atmosphere may also be irregular and chaotic. 

As an evidence of irregular behavior of the model we can see its energy spectrum in \rfg{spectr}.  This spectrum has been calculated from the energy time series of  50 000 days length. We see an uniform amplitude of low frequency  Fourier modes that correspond to periods 1000 --- 50 000 days and  decreasing amplitudes corresponding to periods in a range from 1 to 1000 days. The rate of decrease is linear in logarithmic coordinates with the slope close to $-3/2$. That means the energy of the mode that corresponds to the wavenumber $k$ depends on $k$ as $E(k) \sim k^{-3/2}$.  This power law is close to the power law  of the Kolmogorov's theory of turbulence $E(k) \sim k^{-5/3}$. This fact shows the flow is  turbulent,       
 the solution's behavior is irregular, and there is no dominating  periodic motions.  

\figurecent{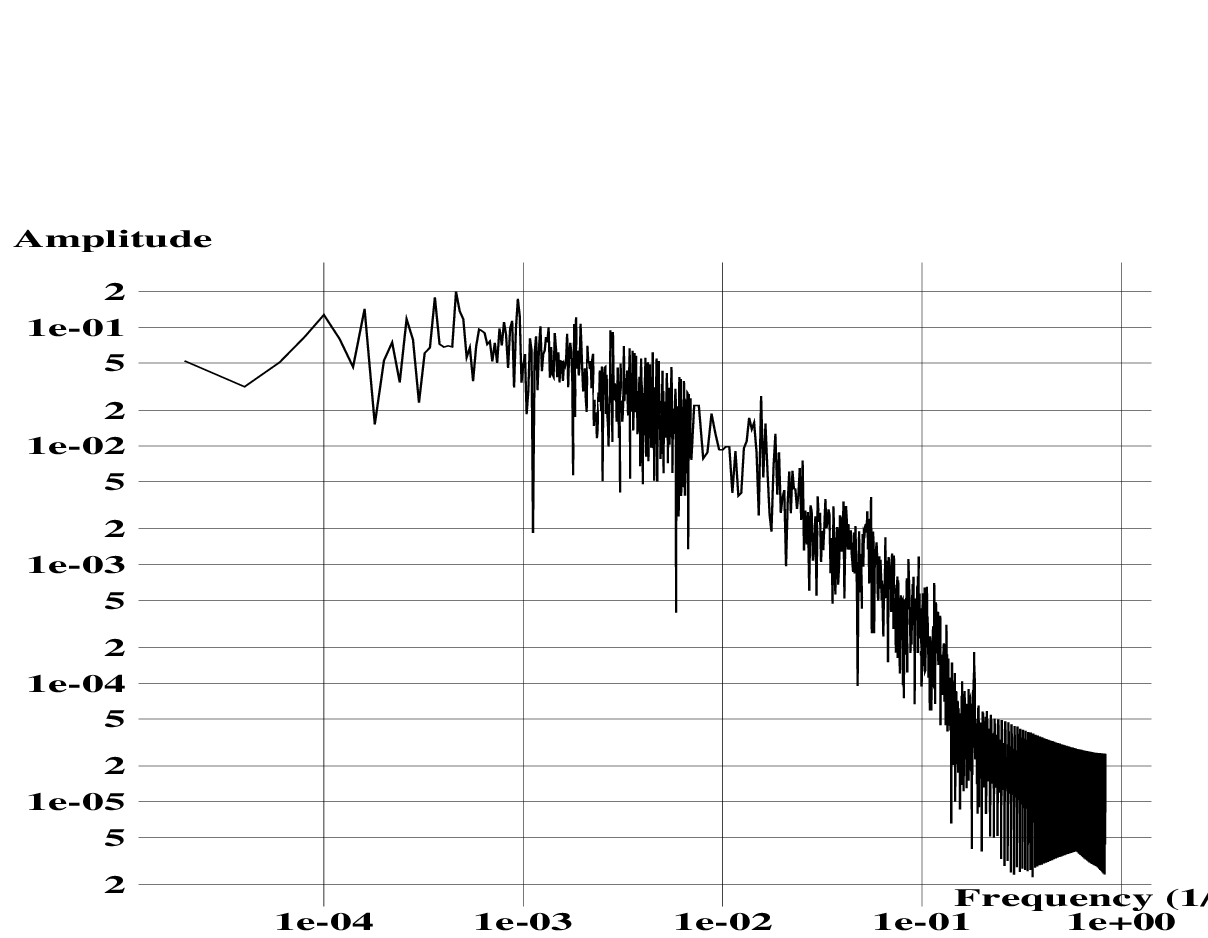}{ Energy spectrum of the model's solution. }{spectr} 

The model was discretised by the  finite element method. 
So far, the  model \rf{btp} under consideration is similar to a barotropic one and the solution produced by the barotropic model of the North Atlantic typically includes a western boundary layer with intense velocity gradients, the advantage of refining the triangulation along the western boundary of the domain is rather clear. A comparison of finite elements (FE) and finite difference (FD) models performed in \cite{LPBB} revealed that the difference arose between simulations by FE and FD techniques can be judged as insignificant when the number of FE nodes is about 6 times lower than the number of FD ones.

The package MODULEF \cite{MODF} has been used to perform a triangulation of a domain. This package produces quasi-regular triangulation of the domain   basing on the prescribed grid nodes on its boundary.  We require the refining of the triangulation near the western boundary and especially in the middle of the domain where velocity gradients are extremely sharps. 
Obtained triangulation is composed of 195 triangles and 436 points. The resolution of this grid is about 40 km near the American coast and about 300 km near the European one.

\section{Results}

In this section we perform several numerical experiments in order to see whether  the procedure is rapid and accurate especially when the observational data are noisy. 

First of all, we shall test the efficiency of the proposed procedure. We generate  artificial ``observational'' data using the same model with the reference topography that is presented in \rfg{topo}B. This fact assures the existence of the absolute minimum of the cost function with vanishing value. Our purpose is to test the minimization procedure and its capacity to  find the topography used to produce  ``observational'' data, assimilating  these data in the model.

The initial guess of the assimilation procedure was taken as  a flat bottom of 4000 meters depth. Thus, we  suppose we have no  preliminary information about the bottom relief in order to perform the assimilation in the most difficult case. During  minimization process, we check two values that characterize the current position of the descent.  The first one is the  value of the cost function \rf{costfun}. It shows the decrease rate of the functional during the minimization.  This value indicates how close is the assimilated trajectory to the reference one, but gives no information about the error in the reconstructed  topography. 

In fact, due to  presence of  null space  of the Hessian of the cost function,  its minimum is not unique. As it has been discussed in \cite{toposens}, the mode of the null space  can be easily seen from a simple analysis of the model \rf{sw}. If we add to the topography $H$ some perturbation which is proportional to $H$ itself $\delta H =\alpha H$, the model remains the same. In this case, only the third equation of the system \rf{sw} is multiplied by $1+\alpha$  and that does not disturb the equality to 0. Hence, the model exhibits no sensitivity to the perturbation $\delta H =\alpha H$ and this mode belongs to the kernel of the operator $G(T)$ \rf{Gmat}. 

So, the assimilation can only help us to find the reference topography multiplied by some arbitrary constant. We must either have some {\it a priori} information about the topography under reconstruction to be able to estimate this constant, or  accept this non unique result of assimilation. In this paper we choose to take into account this arbitrary constant in the post-processing.  So, along with the cost function, we trace also the minimal values of  the difference between the topography on the current iteration $H_n(x,y)$ and the reference topography $\bar H(x,y)$ in form 
\beq
\eta_n =\min\limits_\alpha  \norme{H_n(x,y)-\alpha\bar H(x,y)}  \label{diff}
\eeq
where the norm $\norme{H}$ is defined by \rf{norme}. 

To represent the influence of the bottom relief on the model's trajectory, in \rfg{descente}A we present the enstrophy of two trajectories of  80 days starting from the same initial point that was created as the final point of the  20 years spin-up performed with real bottom topography presented in \rfg{topo}B.  One can see very different behavior of two trajectories. Thus, when we run the model with the same  bottom as during spin-up,  the enstrophy oscillates near its  initial value with no particular tendency. This is natural, because during the spin-up the trajectory has already reached the attractor of the model and remains on it during continuation. On the other hand, when we change the topography of the model, we change its attractor. Hence, the trajectory leaves the former attractor to reach the new one. The enstrophy of the trajectory obtained with flat bottom goes away from the initial point.  

This fact has to be taken into account when the assimilation period is short (shorter than relaxation time to the attractor). In this case we  assimilate  the trajectory going to the attractor,  rather than the model's trajectory on the attractor.

\figureleft{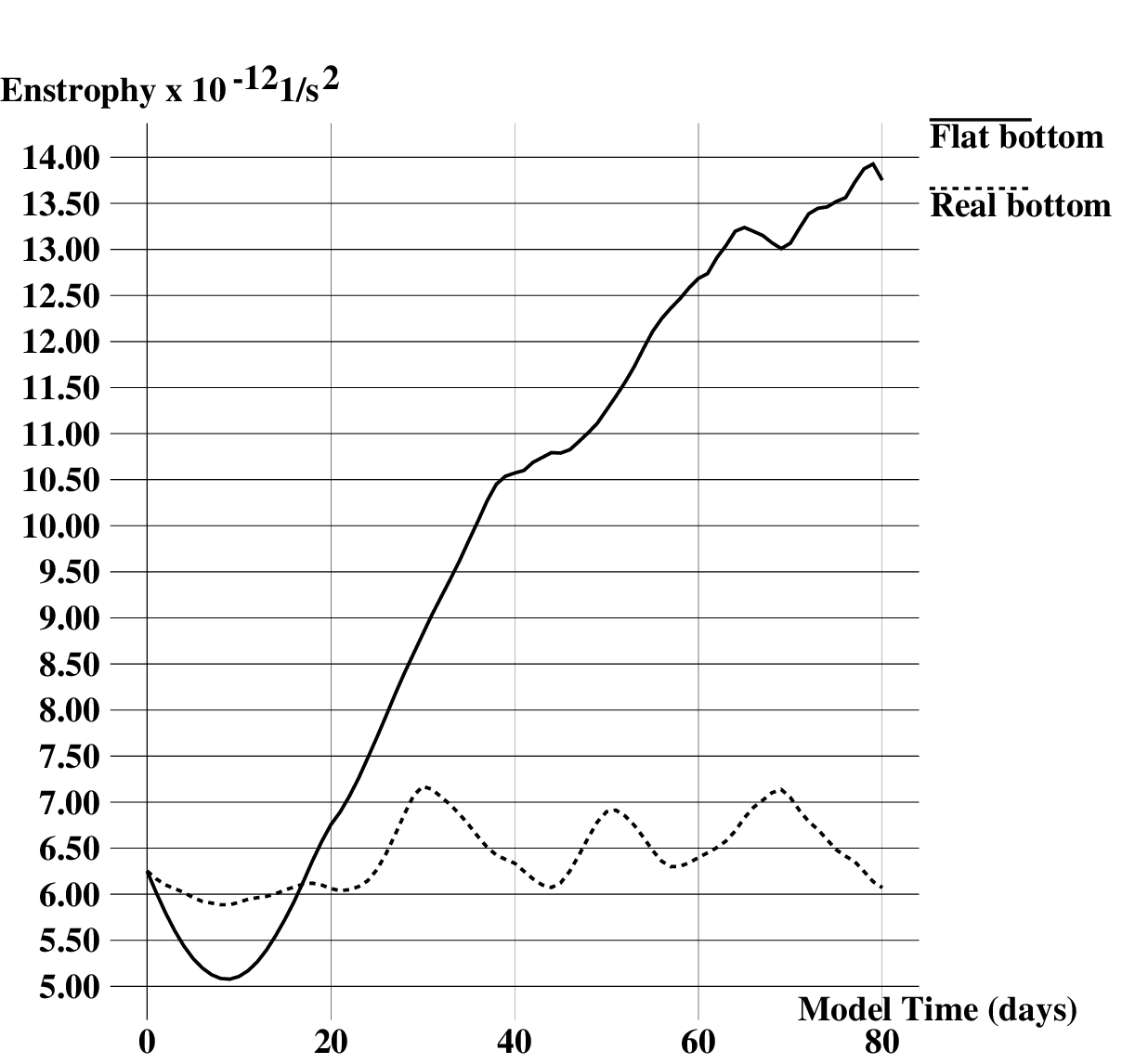}{ Enstrophy of the model with flat bottom  and with assimilated topography} {descente}  
\figureright{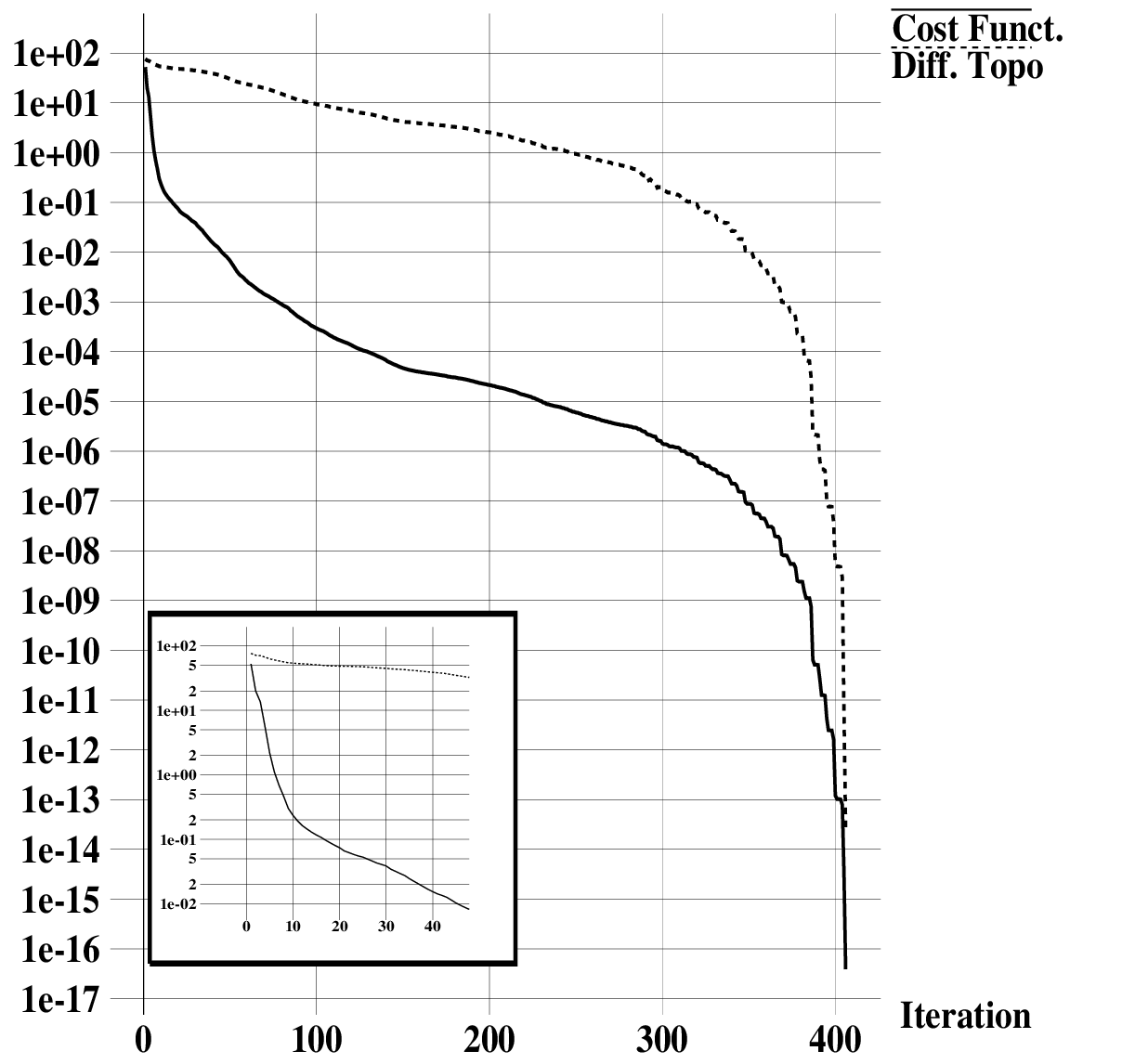}{Cost function  and $\eta_n$ during minimisation }

In the \rfg{descente}B one can see an example of  the evolution of the descent procedure. Starting at  value  $\costfun=54$,  the cost function decreases  rapidly during first 10 iterations. During this time, the contribution of the most sensitive modes of the Hessian is  minimized.  In fact, after 10 iterations the model's trajectory is already close to the reference one. The value of the cost function has been divided by  200. In the same time, the topography is still far from the reference one, the value of $\eta_n$ \rf{diff}  has only been divided by 1.5. The reason of this is simple:  all Hessian modes with low sensitivity have not been damped at this time. These modes  contribute few in the cost function because of low sensitivity of the model. But their contribution in $\eta_n$ is as important  as the contribution of sensitive modes. 

  After that, the  rate of decrease becomes slower.  The procedure needs many iterations to damp modes with low sensitivity.  
  The final value of the minimization vanishes, as supposed, because the cost function in this case has a clear minimum with zero value.

Thus, we may note only few iterations of the assimilation are sufficient to get a good trajectory which is close to the reference one. After 100 iterations the cost function, being divided by $10^5$, is already negligeable. However, the identified bottom topography may not be close to the topography of the reference experiment because of presence of numerous modes with low sensitivity that are not identified at this moment.    

\subsection{Assimilation window: exact ``observations" }

The first question we address concerns the length of the assimilation window $T$.
There is no a priori information for the choice of $T$. In fact, we can choose $T$ as one time step of the model and as one month as well. In order to see the influence of the  window's width on the convergence of the assimilation process, we shall distinguish two aspects of this influence: the window's width itself and the quantity of external information contained in the window.  When we use the cost function \rf{costfun} that is similar to classical  4DVAR cost functions, different windows contain different quantity of the observational information that is  introduced into the model. Each window contains as many observational fields as time steps. 

When we use a 3DVAR cost function $\costfun_3(H(x,y)) =   \norme{\omega(T,H)-\omega_{obs}(T)}^2 $, each window contains exactly the same quantity of information: only one observational field at the end of the window. 
Assimilating data by minimization of the   3DVAR cost function $\costfun_3$ helps us to see the effect of the window's width itself, while  using the 4DVAR cost function we shall see the joint effect of the window  and the information in the window.

When the assimilation window  $T$ increases, the computing time per iteration increases also because the iteration is composed by the integration of the direct model from 0 to  $T$ and backward  integration of the adjoint model. Hence, looking for the optimal $T$, we should  refer to the CPU time  along with  the number of iterations. 

In order to see the dependence of the convergence rate on the window's width, we perform 8 experiments with different assimilation windows $T$. All other parameters are the same in this set. We start from the value $T=0.1$ day, that corresponds to one model's time step. For all subsequent experiments, we double $T$. 

 The ``observations" are produced by the same model, hence, the expected cost function's value must be vanishing at the end of assimilation. We use the criterium  $\costfun(H(x,y)) =10^{-16}$ to stop iterations. 
The number of iterations and the CPU time that are necessary to converge the minimization are shown in the table 1.  The CPU time is expressed in minutes on the Intel Pentium processor. 

\begin{tabular}{|c|c|c|c|c|}
\hline
\multicolumn{5}{|r|}{Table 1. Number of iterations and CPU time necessary to converge  }\\\multicolumn{5}{|r|}{ the minimization process for different $T$. }\\
\hline
    & \multicolumn{2}{|c|}{4DVAR}& \multicolumn{2}{|c|}{3DVAR} \\
\hline    
T(days) & Number of & CPU Time&Number of & CPU Time\\
 &  iterations& & iterations& \\
 \hline
0.1& 2039 & 1.7&  2039    & 1.7 \\
0.2& 1842 & 2.3&  1970    & 2.3 \\
0.4& 1836 & 3.8&  2095    & 4.1 \\
0.8& 1747 & 6.6&  2060    & 7.3 \\
1.6& 1606 & 11.4&  2243    & 15.1 \\
3.2& 1128 & 15.5&  2257    & 29.5 \\
6.4& 891  & 24.2&  2419    & 62.3 \\
12.8&936  & 50.7& 2398    & 123.5 \\
\hline
\end{tabular}

Analyzing the  table 1, we can see several tendencies. 
If we use 4DVAR cost function, the number of iterations becomes smaller  when the assimilation window increases. Hence, each iteration becomes more efficient. It is not surprising, because  it uses more external information. However, smaller number of iterations for longer $T$ can not compensate increasing of the CPU time per iteration. Each iteration requires almost double CPU time for double $T$, while the number of iterations is far from being divided by 2. And, consequently,  CPU time is lower for low assimilation windows. 

In the case when the 3DVAR cost function is used,  the number of iterations seems to be constant for  windows smaller than 1 day, but when $T$  increases, the minimization require slightly more iterations to converge. Indeed, no additional information is introduced into the model by enlarging of the window. Therefore, there is no hope  for the assimilation to be more efficient. In this case, hence,  the convergence is undoubtedly more rapid when $T$ is small.

As a consequence, when  observations contain exactly the necessary information,  the best assimilation window is one time step for both 3DVAR and 4DVAR (they coincide in this case).

\subsection{Assimilation window: noisy ``observations" }

However, this experiment was carried out in frames of  idealized situation.  Artificial observations created by the same model contain the exact information about the model's topography. It is not the case when we use real observations containing errors of measurements. In addition to this, real observations are produced by many different physical processes that are not taken into account in the model.   In this 
case, the information about topography contained in observations  can no longer be considered as exact.  

In order to simulate the influence of possible errors in observational fields we add a white noise of several different  amplitudes  to  assimilated data. 

An example of this assimilation experiment is presented in \rfg{bruit}. As above, in order to produce the data to be assimilated, we run the model with the topography $\bar H(x,y)$ \rfg{topo}A.  After that, a white noise has been  added to these data at each grid point and at each time step: 
\beq
\omega'(x,y,t)= 
\omega(x,y,t)+\varepsilon
\fr{\norme{\omega(x,y,t)}}{\norme{ r(x,y,t)}}\tm r(x,y,t) \label{noise} 
\eeq 
where $r(x,y,t)$ is a random real number from the interval $-0.5 \ldots 0.5$.

 An example of the noise effect  is shown   in \rfg{bruit}A. The solid line represents the enstrophy of the reference experiment. Irregular dashed line with small dashes shows the trajectory with  noise. So far the noise is  irregular, it produces additional enstrophy to flow fields, providing the dashed line  is almost always above the solid one. The smooth dashed line with long dashes represents the enstrophy of the trajectory of the model with identified topography. One can see this line is almost indistinguishable from the reference one.  

The convergence of the cost function and the difference $\eta_n $ defined by \rf{diff} during the assimilation process are shown in \rfg{bruit}B. Contrary to \rfg{descente}B, neither the cost function nor the difference vanish at the end of assimilation. They both tend to some final error. It is clear this final error is  due to the noise that can not be assimilated by the model. 

\figureleft{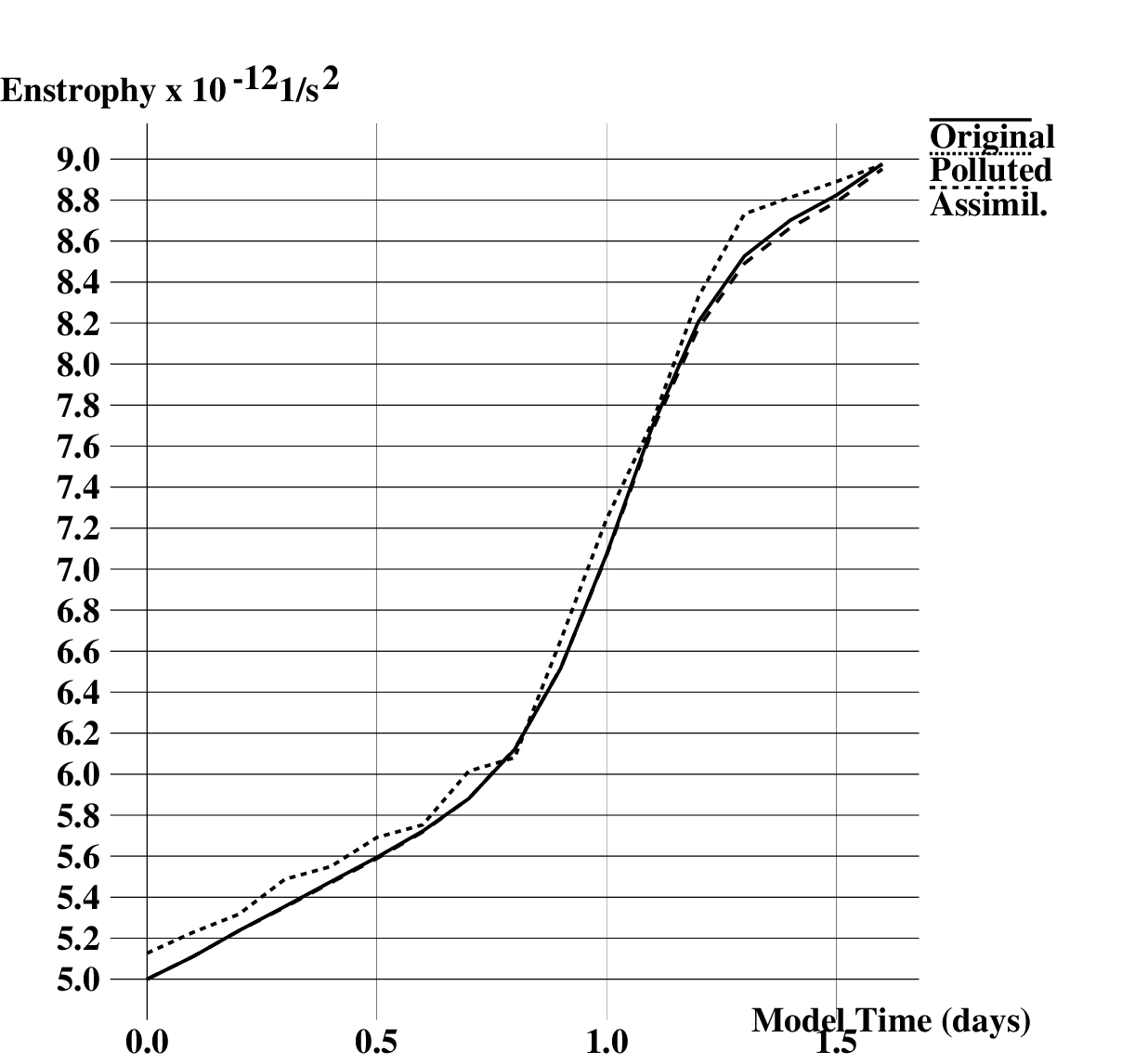}{ Enstrophy of the reference (solid line),  noisy (short dashes)  trajectories and the assimilation result (long dashes)} {bruit}  
\figureright{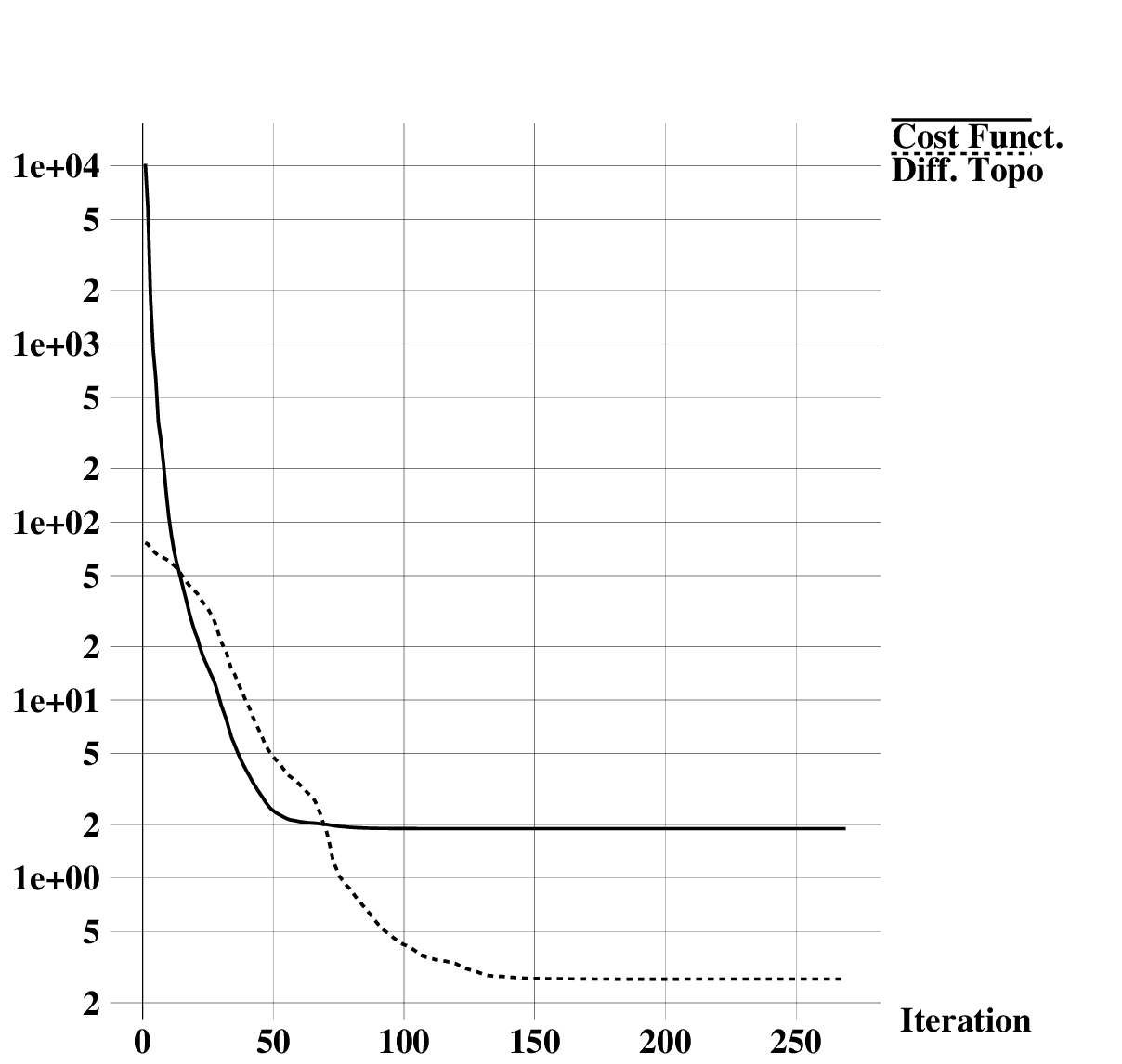}{Convergence of the cost function (solid line) and of the difference $\eta_n $  (dashed line) in assimilation of noisy data.}

The question we can ask, whether small values of $T$ are still optimal in the case of assimilation of  noisy data. In fact, in the case of artificial "observational" data generated by the same model without noise, the result of identification of the topography was always the same. We were always able to reconstruct the exact topography of the model and reduce the cost function's value  to zero.  We were optimizing the computational cost rather than the result of assimilation. But
when we assume the presence of noise in  observational  data, the optimality  condition is not the same as in the previous case. Now it is better to obtain more accurate topography than low computing time.  It is not the convergence rate that we want to optimize, but the final error in the reconstructed topography.

In order to see whether there exists an optimal  assimilation window in this case, we use noisy data $\omega'(x,y,t)$  as artificial observations in the assimilation procedure.  A set of experiments has been carried out  with different  windows. In the first experiment the window length was taken as the smallest possible   one time step window. In each next experiment, the window length was two times longer. $T=0.1 ,0.2, \ldots 51.2$ days. Several examples of the convergence of $\eta_n$ are shown in \rfg{t-b}A for $\varepsilon=10^{-3}$. We see in this figure, when the noise is present, small assimilation windows  provide worse results. Despite the convergence rate remains better for small $T$, the final error of reconstruction is relatively big. Thus, if assimilation window is restricted to $T=0.1$ days, the value $\eta$ converges rapidly to the value $2.8\tm 10^{-1}$ and after that remains stable. Multiplying  $T$ by 4 makes the convergence slower, but allows to reach $\eta = 1.7\tm 10^{-1}$.  Using longer assimilation windows allows us to reduce final residual error in topography to $\eta = 2\tm 10^{-2}$ with T = 3.2 days; to $\eta = 9.4\tm 10^{-4}$
 with $T = 25.6$ days and even to $\eta = 4.4\tm 10^{-4}$ with $T = 51.2$ days. However,
 the convergence becomes slower for longer windows. Not only the number
 of iterations of the descent procedure increases, but each iteration requires
 more computer time, because on each iteration the model and its adjoint are
 integrated for a longer time.

There exists, hence, no optimal value of $T$. Longer windows provide always better accuracy of the identified topography. This is clearly seen at \rfg{t-b}B, where the final value of  $\eta$ is plotted for each window $T$ for three different amplitudes of added noise $\varepsilon=10^{-4}, 10^{-3},10^{-2}$ (these amplitudes are shown
> in the figure as small, medium and big noises respectively).

\figureleft{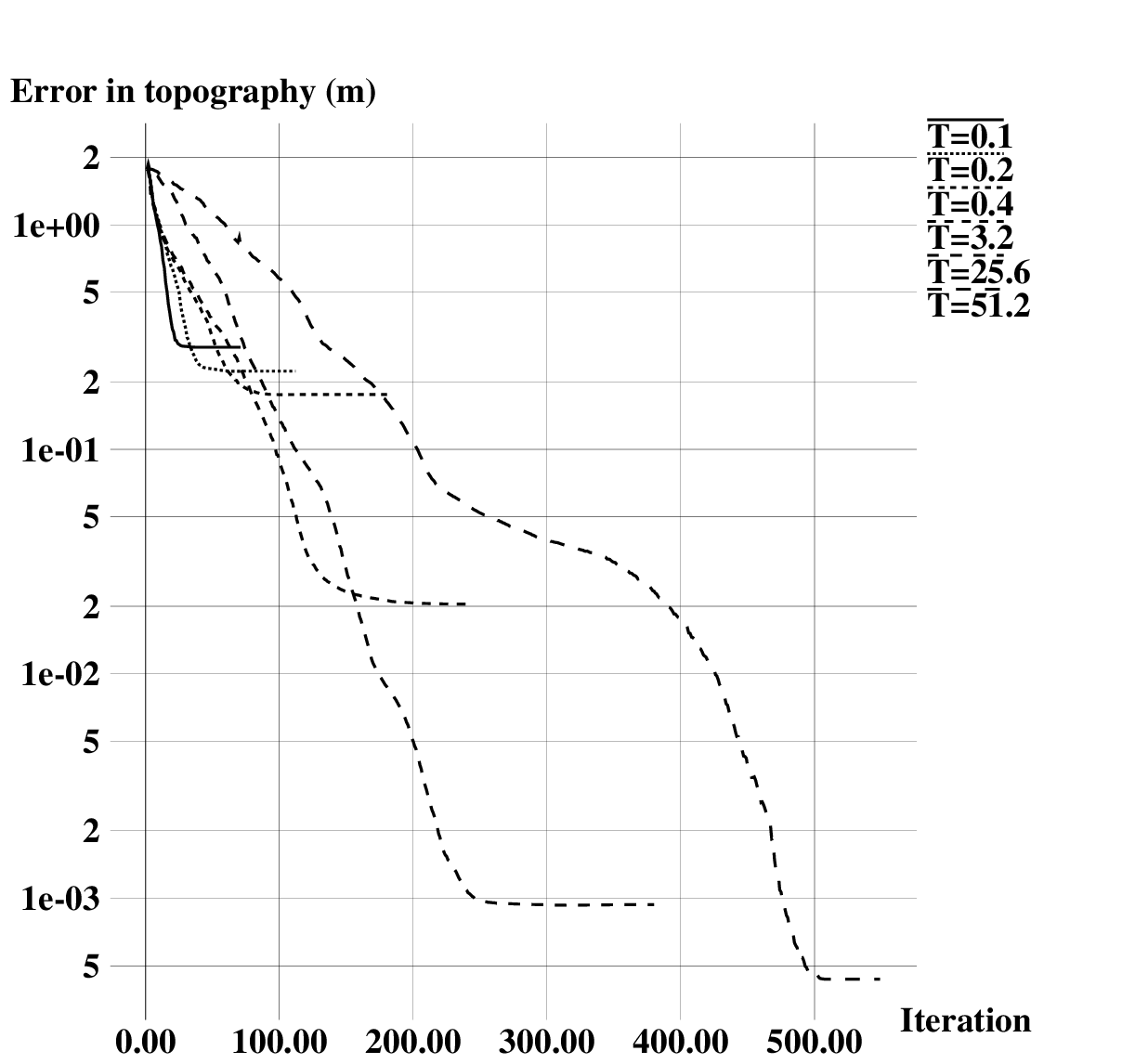}{ Convergence of $\eta_n $  for different assimilation windows $T$ in the case of noisy data.} {t-b}  
\figureright{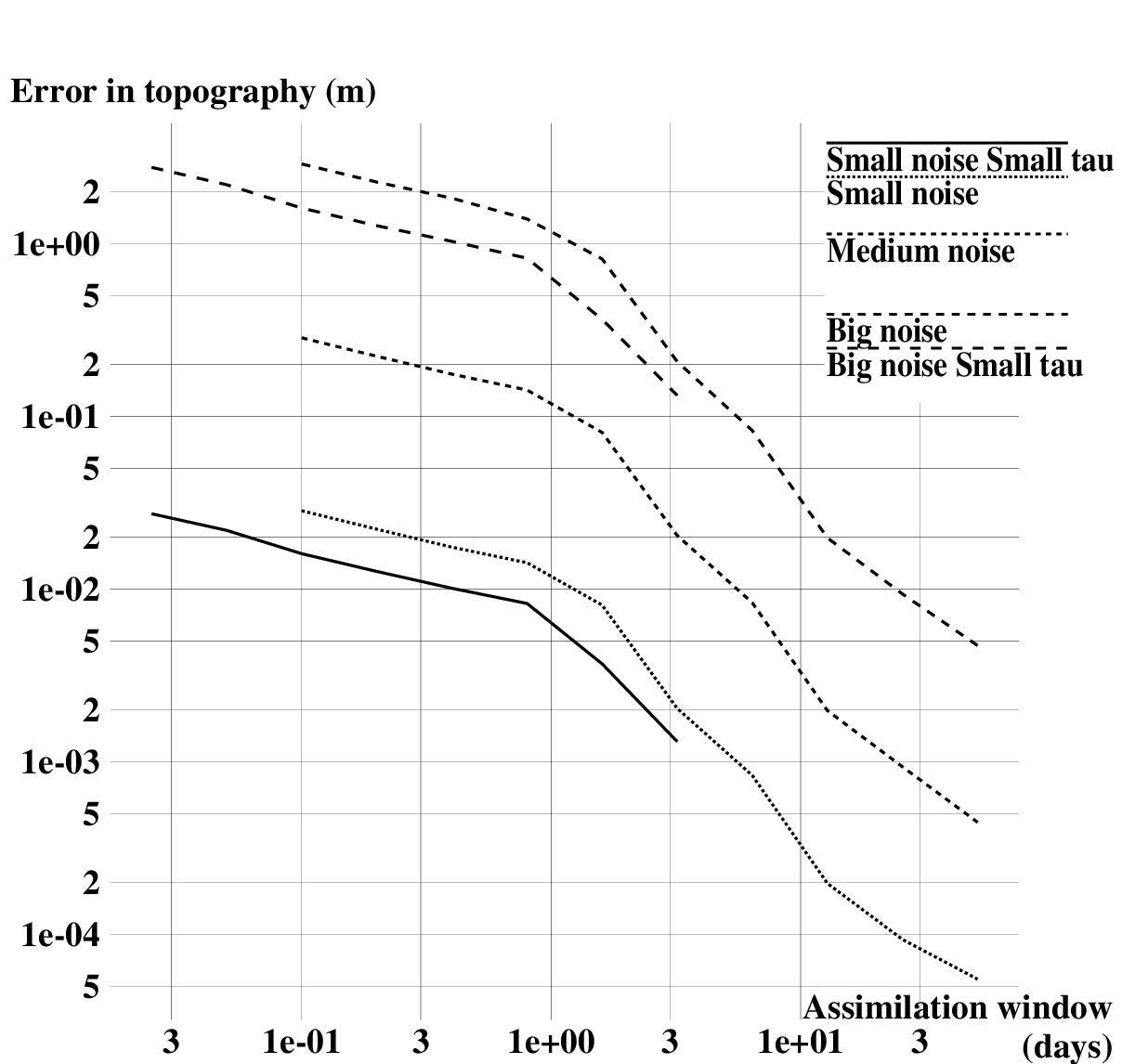}{Final value of $\eta_n $ for different assimilation windows $T$ with small (solid lines), medium (short dashes) and big (long dashes) noises. }

  Increasing  assimilation window, we increase the quantity of assimilated information and improve the accuracy of the resulting topography. This fact seems to be natural if we take into account that assimilated data are noisy. More noisy information introduced into the model helps to reduce the noise impact and results in better accuracy. This is similar to  elementary statistical notion: if we calculate an average of noisy data, we get better accuracy with more data. 
  
   Thus, for example, if we calculate averages of $N$ random numbers distributed in the interval from -1 to 1 we can hope to obtain zero for infinite $N$. For any finite $N$, the average will be different from zero, but its mean deviation from zero will be smaller for bigger $N$.
   In this simple example, the mean deviation of the average from 0 is proportional to the square root of the  mean variance. The dependence of the mean variance on $N$ is determined by the central limit theorem. The mean variance of averages of $N$ random numbers is inversely proportional to $N$. The mean deviation, hence,  will decrease as $N^{-1/2}$. 

This is not  the case in experiments with different assimilation windows. So far, the assimilation we use is similar to 4D-Var, the quantity of noisy external information introduced into the model is proportional to $T$. Therefore, we can hope the decrease rate of residual error $\eta$ will be proportional to $T^{-1/2}$. However, in \rfg{t-b}B we see two different decrease rates. 

The first   one can be seen for windows in range from 0.1 to 2 days. Indeed, the value of $\eta$  decreases as $T^{-1/2}$ in this range. This fact shows the statistical law dominates for these windows. The noise contained in data influences  directly  assimilation results and we obtain the dependence on $T$ predicted by the central limit theorem. 

 But when the window is longer  than 2 days, we see more rapid decrease of residual error in topography. On this part the decrease rate is close to $T^{-3/2}$ and we can deduce the influence of noise in data is not as straightforward as before. In order to understand whether the change of the decrease rate has a physical meaning and occurs effectively at assimilation window $T=1-2$ days, or this change has numerical origins and corresponds to several  time steps of the model rather than to physical time, we perform two similar experiments with 8 times smaller time step. Assimilation windows in these two experiments were chosen in range from 0.025 days (one small time step)  to 1.6 days (128 small time steps). Corresponding lines can also be seen in \rfg{t-b}B.  Comparing them with lines corresponding to the 0.1 days  time step, we see the same decrease rate:  $T^{-1/2}$ for small windows and acceleration for bigger windows. That means the change in decrease rate does not related to numerical time stepping and must have physical origins. We see also both lines with smaller time step lie below corresponding lines with normal time step. It is natural: so far, the time step is smaller, more external information is introduced into the model each time unit. And more assimilated information results in better precision.  
 
 Moreover,  comparing lines with small and normal time steps, we can see  the residual error in topography depends on the quantity of assimilated data only. No matter how long was the assimilation window, it is the quantity of time steps (which is equal to the quantity of assimilated observational fields) that determines the residual error. Residual errors obtained with one, two or three time steps are approximately equal with both the time step length 0.025 and 0.1 days.

Hence, the change in decrease rate has some physical reason.  
 Indeed, it was shown in \cite{toposens} the time scale 2-3 days is important in  the sensitivity analysis   of this model.  On time scales lower than 3 days, the sensitivity is linear in time. That means an error in the  model's solution induced by  a topography perturbation grows linearly in time. But, on  longer time scales, error growth  becomes exponential. That means, during 2-3 days of  model time the perturbation is introduced into the model and, after that, it follows the model's dynamics. During the phase of introduction,  linear transition of perturbation from topography to model's variable dominates, resulting in  linear  dependence on time. But after 2-3 days, it is the model's  dynamics that governs the error evolution. Being non-linear and intrinsicly unstable, the dynamics ensures exponential error growth  of a perturbation. On these time scales, topography perturbation evolves like any other perturbation from any other source. 
 
Assimilating noisy data we see a similar  difference between small and large scales. 
If the model's dynamics  has not enough time to get adapted to the noise contained in assimilated data then the noise effect is directly transmitted to identified topography.   So, its influence on assimilation's results is similar to influence of a pure white noise. On large time scales the noise is transmitted by the dynamics and modified during transmission.  As a result, we have more rapid decrease of the residual error.

\subsection{Noise amplitude and assimilation error }

In order to quantify the relationship between  noise in model's parameters and  error in the reconstructed topography, we perform a set of experiments  assimilating  noisy data. We examine the amplitude of the noise and also it's origins. It is clear, bigger noise will provide bigger residual error in topography. But the source of the noise may also be important.

 First, we add the noise in  the model's initial conditions  simulating the influence of  interpolation or residual errors of reconstruction of initial point in the data assimilation process. 
Second, we perturb the forcing ${\cal F}(x,y)$ of the model \rf{btp} in order to simulate the difference between parametrisation of physical processes in the model and  real physical processes in observational data. And third, we add the  noise to all grid-points at all time steps of the artificial observations to simulate the measurements errors. 

In the third case we distinguish two situations. In the first one,  the noise is added directly to the vorticity $\omega$, which is supposed to be an observable variable and no transformation is necessary to get the model's variable. 
In the second situation, we add the noise to the sea surface height because it is the altimetry that is the most probable candidate to be an observable variable in real data assimilations. 

To construct the relationship between the model's variable and altimetry we can use the geostrophic assumption. If the gradient of the sea surface elevation $h$ is in equilibrium with  the Coriolis force, we can write
$$ -fv = -g\der{h}{x}, \;\;\;\; fu=-g\der{h}{y}$$ 
Using the expression \rf{strf-uv} we get the sea surface elevation $h$ is proportional to the streamfunction:
\beq 
\psi = \fr{gH}{f } h
\eeq
Taking into account the equation \rf{stat} we get 
\beq
\omega = g \der{}{x} \fr{1}{H} \der{}{x}\fr{hH}{f } + g\der{}{y} \fr{1}{H}\der{}{y}\fr{h H}{f } \label{altimet}
\eeq
If the depth $H$ and Coriolis parameter $f$ are always positive, the operator of the equation \rf{altimet} is invertible and for any $\omega$ we can find  corresponding $h$. Thus, if the artificially generated ``observations" are not polluted by noise, then there is no difference between assimilation of vorticity data and assimilation of the sea surface height. Both of them contain  the same exact information about topography.  Performing the reference experiment we can generate $h$ as observable variable instead of vorticity using the inverse of the equation \rf{altimet}. But assimilating these data we have to obtain  the model's variable from observations, i.e. to apply direct operator \rf{altimet} to the  altimetry.  

However, when noise is present in data, assimilation's results  may be different in two cases. If noise is added to the sea surface height, the following transformation of $h$ to $\omega$ will also transform the noise.  Vorticity, in this case, will be polluted by  the application of the operator \rf{altimet} to the noise, rather than the noise itself.  

In order to see the influence of  the noise transformed by the operator \rf{altimet}, we perform  the fourth experiment in which the similar random  noise has been added to the  sea surface height in each grid point and  at each time step.  The model's variable, vorticity, was obtained from noisy $h$ by formula \rf{altimet}.

Before  data assimilation and topography identification, we shall see the influence of  different perturbation's sources on the model's variable. 
Each of  above mentioned parameters (initial conditions, forcing, vorticity and SSH) were perturbed  by the  noise \rf{noise} of a prescribed amplitude $\varepsilon=10^{-3}$. In \rfg{assvar}A we trace  evolution of the norm of the difference between the reference trajectory $\omega$ with no noise and noisy trajectory $\omega'$ during 150 days model run:
\beq
\xi(t)=\norme{\omega'(x,y,t)-\omega(x,y,t)}=\sqrt{\int (\omega'(x,y,t)-\omega(x,y,t))^2 dx dy} \label{xi}
\eeq
Evolution of $\xi$ during first 15 days is presented as a zoom in this figure.

The solid line in this figure represent the evolution of the noise  added to initial conditions: 
$$\omega'_0(x,y)= \omega_0(x,y)+\varepsilon
\fr{\norme{ \omega_0(x,y)}}{\norme{r(x,y)}}\tm r(x,y) $$ 
where $r(x,y)$ is a random real number from the interval $-0.5 \ldots 0.5$ . 

Value $\varepsilon=10^{-3}$ signifies the norm of initial point is perturbed by 0.1\% 
of its value: 
$$\fr{\norme{ \omega'_0(x,y)-\omega_0(x,y)}}{\norme{\omega_0(x,y)}} =10^{-3}$$
This line starts at $\xi=9\tm 10^{-6} \fr{1}{s}$ and  grow rapidly during first 4 days. This is the case of well known super-exponential error growth (see, for example, \cite{Nicolis95}) that reduces short-range  predictability of chaotic systems because local Lyapunov exponents (that govern the error growth on short time scales) are bigger than global exponents on long time scales \cite{kaz99}. 
After four days period, the value of $\xi$ grows exponentially (linearly in logarithmic coordinates), with growth rate determined by long-time Lyapunov exponents. Beyond the 150 days interval, the line will reach  saturation with values  about   $\xi= 10^{-2} \fr{1}{s}$ which represents the characteristic radius of the model's attractor.   

The dotted line in \rfg{assvar}A represents the evolution of $\xi$ in the case when the noise has been added to the forcing of the model. This line, obviously,  starts from 0 but the rapid growth lasts more than 20 days. This happens because modification of the forcing of the model changes it's attractor. The trajectory evolves first toward the new attractor and after that, the value of $\xi$ grow also exponentially, in a similar way as  the solid line. 

Two oscillating lines represent the noise added  to all grid-points at all time steps to the reference trajectory in order to simulate the measurements errors. This noise is not governed  by the model's dynamics, consequently these lines are oscillating about constant values. The lower oscillating line represents the noise added to the vorticity directly. This line oscillates about $9\tm 10^{-6} \fr{1}{s}$, the same value from which  starts the solid line. This is natural because the noise that has been added to initial point is the same as the noise added to the vorticity.  The upper line is the representation of the noise in the sea surface height. The difference in position of these two lines shows that  transformation of the noise by operator  \rf{altimet} increases the amplitude of the noise approximately two times. Hence, adding noise in the SSH, we should expect bigger  residual error in the reconstructed topography than adding noise in the vorticity directly. 

And finally, in order to compare the influence of noise in all these parameters with the influence of the same noise in the bottom topography of the model, we plot the fifth line. The  dashed   line in this figure represents the evolution of the norme $\xi$ \rf{xi} when the noise of the same amplitude $\varepsilon$ has been added to the topography. The amplitude $\varepsilon=10^{-3}$ means the depth was perturbed by 4 meters in average. 
This line reveals stronger influence of  the noise in topography field on the trajectory. The value $\xi$ is equal to zero at the beginning
 of integration as well as in experiment when noise is added to the forcing of the model. Also similarly to the experiment with forcing, rapid increase of $\xi$ lasts approximately 20 days. As well as the forcing modification,  modification of topography also changes the model's attractor and the trajectory goes  also first to the new attractor. However, the noise in topography generates much higher  growth rate of $\xi$  than  the noise in forcing. That means, the attractor is more modified by a little change of topography than by an equal  change of the forcing. 
This fact shows the importance of better identification of the topography of the model.

\figureleft{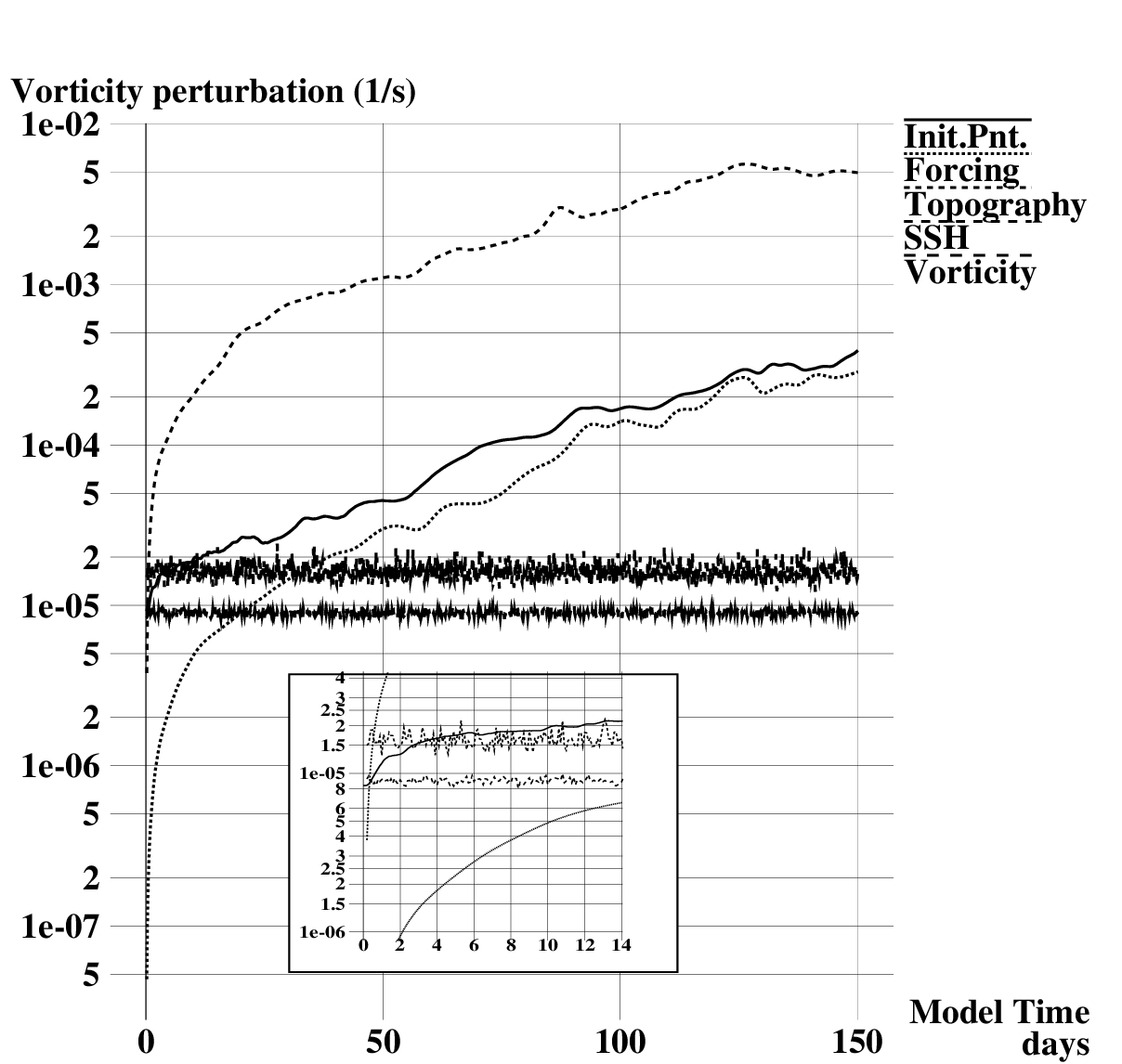}{ Evolution of  vorticity perturbation for different sources: noise in initial point (solid line), noise in forcing (dotted line), noise added to vorticity and SSH (oscillating lines) and noise in the bottom topography (dashed line).} {assvar} 
\figureright{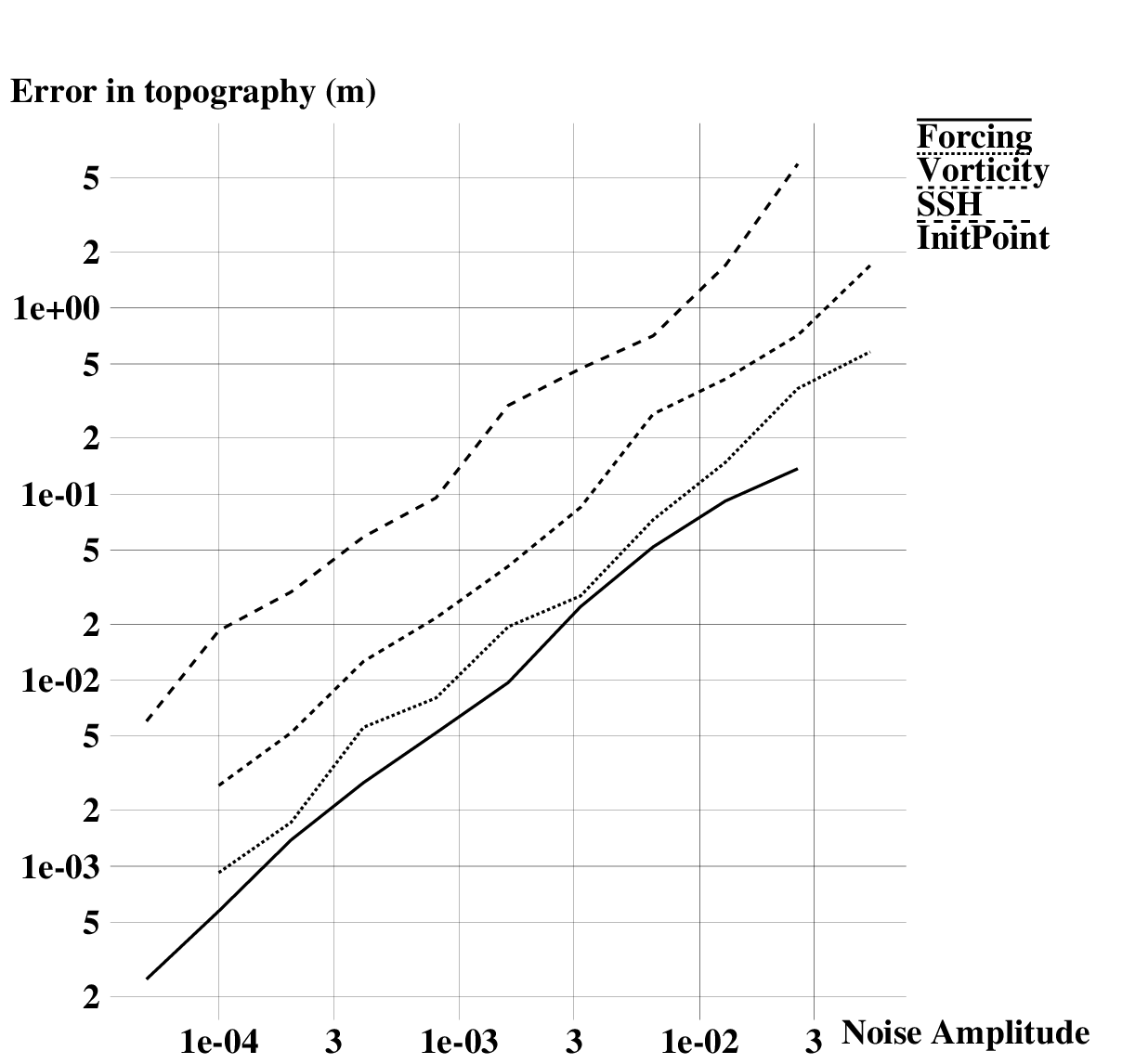}{Residual error in topography $\eta$  as a function of  noise amplitude $\varepsilon$.  }

Error obtained in the reconstructed topography for different perturbations in model parameters and its trajectory is shown in the \rfg{assvar}B. 
The experiment was carried out in four sets. In each set one and only one  parameter (initial point, forcing, vorticity or SSH of the  trajectory itself) has been perturbed by random noise with ten different amplitudes. In the first set the ``observational" data were generated by the model with perturbed initial conditions.  The perturbation amplitude was doubled in each of ten experiments beginning at the amplitude $\varepsilon = 5\tm 10^{-5}$. The amplitude of perturbation in the tenth experiment was equal to $\varepsilon = 2^9\tm 5\tm 10^{-5}=2.56\tm 10^{-2}$.   

The second set was performed with ``observational" data obtained with non-perturbed initial point, but with  noisy forcing. As before, the noise amplitude was doubled in each experiment beginning at the same value $\varepsilon$. And in the third set, neither initial point, nor forcing were perturbed, but the resulting trajectory was subjected to perturbation with amplitudes beginning at $\varepsilon =10^{-4}$. In the fourth set,  noisy sea surface height was used as observable variable with the same noise amplitudes as in the third set.

 Assimilation window in all  these experiments was chosen to be $T=5$ days. The topography shown in \rfg{topo}A used to create the reference ``observational" data was the same in all experiments . 
 The data assimilation of noisy data was performed up to stabilization of the minimization processed.   We  trace then the resulting norm of the difference between the reconstructed topography and the exact one, used to create the reference ``observational" data.

The dashed  upper line with long dashes shows the dependence of the final error in topography on the amplitude of the perturbation of initial conditions. Higher position of this line indicates   stronger influence of  errors in  approximation of initial point of the model. This is not the case when the forcing is perturbed (lowest solid line). We can see that the model exhibits lower sensitivity. The amplitude of the noise in the forcing can be 100 times higher but the assimilation provides equal error obtained in the reconstructed topography. 

The influence of  noise in the trajectory of the model (dotted  line corresponds to the case with noisy vorticity and dashed line with small dashes corresponds to perturbation of the sea surface height) is a little higher than the influence of noise in the forcing but lower than the influence of noise in initial point. Adding noise in the sea surface height results in a bigger error in reconstructed topography. This is the consequence of already mentioned fact:  the operator \rf{altimet} applied to a random noise results in a noise of bigger amplitude.

The slope of all lines in \rfg{assvar}B is equal to one. That means the value of $\eta$ \rf{diff} is linear function of $\varepsilon$.

\subsection{Beyond the assimilation window }

An important difference between identification of optimal initial point of the model and its optimal topography consists in the fact that topography is an internal model's parameter. It must be identified once for all model runs,  while initial conditions are external parameters and must be identified for each particular model run. The question we should ask in this case, whether the  topography identified once by data assimilation is valid for other model runs that start from other initial conditions and are situated in other parts of the model's attractor. 

In this experiment we perform different model runs using results of assimilation of noisy data obtained in previous section. All runs start from different arbitrary  points on the attractor of the model with the reference topography. 
These initial points have been chosen as arbitrary points on the long trajectory of the model integrated with the reference topography.

As we have seen, assimilation of noisy observations results in an error of topography reconstruction. Consequently,  it is hopeless to obtain the same solution with two different topographies. So, we shall analyze the evolution of the relative difference between  trajectories of  models with the reference topography and with the reconstructed topography $H_{reconsr}$ containing errors due to assimilation of noisy data: 
\beq
r^2_{reconsr}(t)=\fr{\int_{\Omega}\biggl(\omega_{H_{ref}}(x,y,t)-\omega_{H_{reconsr}}(x,y,t)\biggr)^2 dx dy}{\int_{\Omega}\biggl(\omega_{H_{ref}}(x,y,t)\biggr)^2 dx dy}
\eeq

This difference is compared with the effect of a simple perturbation of topography by  random values at each grid-point: 
$$H_{prtrb}(x,y)= H_{ref}(x,y)+\varepsilon
\fr{\norme{ H_{ref}(x,y)}}{\norme{r(x,y)}}\tm r(x,y)  $$
  The amplitude of perturbation in each experiment was exactly equal to the relative residual error in the topography after assimilation. The model with perturbed topography  has been integrated from the same initial point as the reference model for 180 days. During this time  we calculate also the relative difference:
\beq
r^2_{prtrb}(t)=\fr{\int_{\Omega}\biggl(\omega_{H_{ref}}(x,y,t)-\omega_{H_{prtrb}}(x,y,t)\biggr)^2 dx dy}{\int_{\Omega}\biggl(\omega_{H_{ref}}(x,y,t)\biggr)^2 dx dy}
\eeq

Evolution of these relative differences $r_{prtrb}(t)$ and $r_{reconsr}(t)$ 
is shown in \rfg{afterassim}A.  
Ten values of relative residual errors in topography 
\beq
\zeta =\fr{\eta }{ \norme{ H_{ref}(x,y)} }=\fr{\min\limits_\alpha  \norme{H(x,y)-\alpha H_{ref}(x,y)}}{ \norme{ H_{ref}(x,y)} }
\eeq
in range from $2\tm 10^{-5}$ to $0.01$ have been tested together with ten  equal perturbation amplitudes $\varepsilon$. So far the average ocean depth is close to 4100 meters, this range corresponds approximately to errors in physical depth $\eta$ from $8\; cm$ to $41\; m$.   

Evolution of  $r_{prtrb}(t)$ and $r_{reconsr}(t)$  corresponding to the lowest and to the highest $\eta$ are shown in \rfg{afterassim}A.  Two solid lines represent the difference $r_{reconsr}(t)$ for $\zeta=2\tm 10^{-5}$ and $0.01$. Two dashed lines  show the difference $r_{prtrb}(t)$ for the same values of $\varepsilon$. One can see that for both $\eta$ the difference obtained with the model with reconstructed topography is 5-10 times lower than with randomly perturbed topography. Despite  amplitudes of the random perturbation and of the residual error are equal to each other,  the trajectory of the model on the reconstructed topography corresponds better to the reference model.  This can be explained by the fact, that all modes in the topography  the model's solution is sensitive to, have been damped during the data assimilation procedure. The residual error is essentially concentrated in modes that provide low impact on the trajectory. While the random perturbation of the topography adds noise  to all modes and results in bigger model error. 

\figureleft{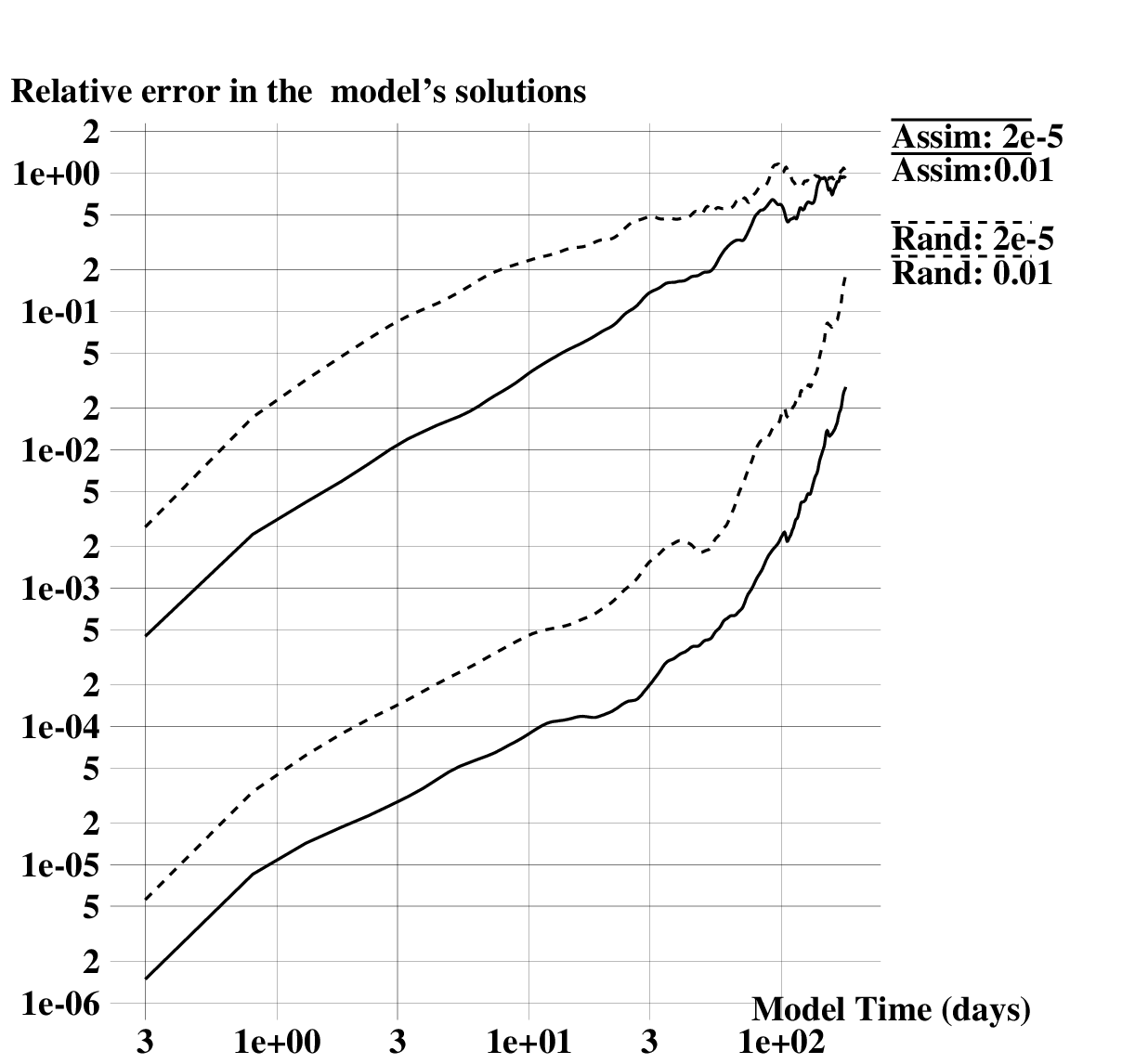}{ Evolution of the relative difference $r(t)$ between the reference model and models with inaccurate topographies: reconstructed topographies (solid lines) and randomly perturbed topographies (dashed lines).  } {afterassim} 
\figureright{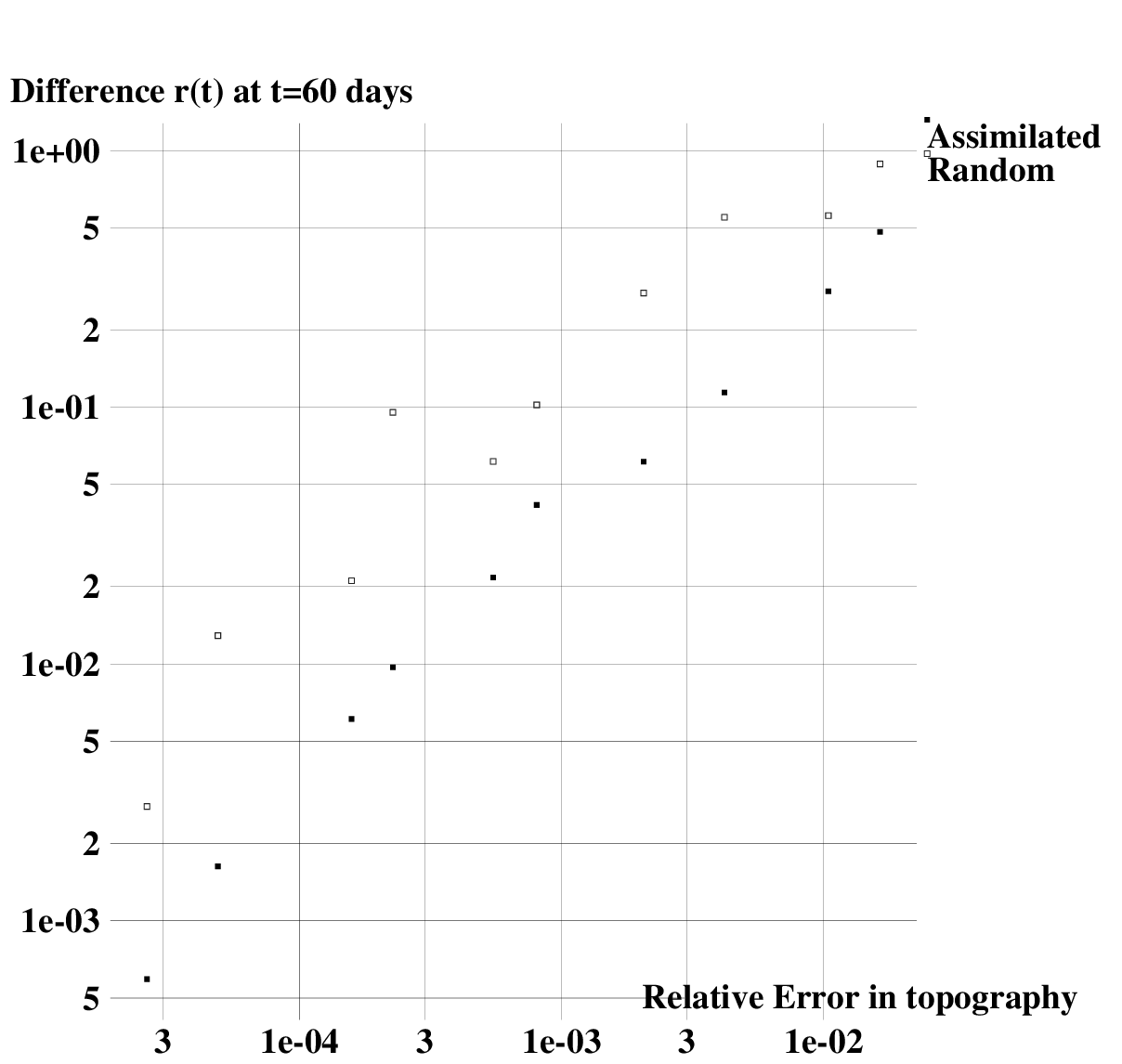}{ Relative difference $r(t)$ at 60th day versus relative error in topography $\zeta$ or noise amplitude $\varepsilon$.   }

We can see the same feature in \rfg{afterassim}B where  differences $r_{reconsr}(t)$ and  $r_{prtrb}(t)$ are shown  at time $t=60$ days for all ten amplitudes of the residual error and of the random perturbation.  Model with the reconstructed topography shows 3-10 times lower difference with the reference model than the model with  randomly perturbed topography. The dependence of the difference at  $t=60$ days on the amplitude of residual error is linear in logarithmic coordinates with the slope equal to 1. That means this dependence is simply linear.  If we remind that the dependence of the residual error on the noise amplitude in assimilated data is also linear, we can deduce that the error  in  the model trajectory is proportional to the error in assimilated data. 

\section{Conclusion}

We have studied the procedure of data assimilation for identification of the topography  for a simple barotropic ocean model. Comparing this procedure with now well developed data assimilation intended to identify optimal initial data, we can say there are both common points and differences as well. 

Tangent  and  adjoint model   are composed by two  terms $A$ and $B$ (\rf{A}, \rf{B}).  The first one, $A$, governs   the evolution of a small perturbation by the model's dynamics. This term is common for any data assimilation no matter what parameter we want to identify.  The second one, $B$,  determines the way how the uncertainty is introduced into the model. This term is specific to the particular variable under identification. This term is absent when the goal is to identify initial point because the uncertainty is introduced only once, at the beginning of the model integration.

The presence of null space of the sensitivity operator constitutes another particularity of the data assimilation in this case. Exact topography can not be reconstructed because there exists a mode the model is not sensitive to. Adding this mode to the topography of the model does not change its solution. Consequently, this mode can not be identified from the model's trajectory. We must either have some a priori information about topography, or modify the model in order to suppress the null space. In this paper  the null space can be suppressed by replacement of the average ocean depth $H_0$ by the real depth in each point $H(x,y)$ in the forcing term $\frac{{\cal F}(x,y) }{\rho_0 H_0} $ in \rf{sw}. The presence of one-dimensional  null space has also been pointed out in \cite{LoschWunsch} where the bottom topography was used as a control parameter for a steady state solution of a linear shallow-water model in a zonal channel. However, authors note the null space vector shows a nearly pure grid-scale noise that can be explained by technical aspects of the numerical scheme on a C grid. They point out this issue can be avoided by choosing the depth at the velocity points as control parameter.

When  assimilated data are perfect and contain exact information, the minimization procedure always converges to the reference topography. In this case, it is preferable to use shorter assimilation windows because they require less CPU time due to shorter integration of direct and adjoint models on each iteration. The number of iterations necessary to assimilate the data depends on the quantity of these data only. When 3D-VAR assimilation is used, almost the same number of iterations is necessary to converge in experiments with different assimilation windows because  the quantity of external information introduced into the model is the same for any assimilation window in 3D-VAR: just one field at the end of the window. And the same quantity of information leads to the same quantity of iterations. 

When the 4D-VAR assimilation is used, new information is introduced at each model's time step. The quantity of external information is, hence, proportional to the length of the assimilation window. And in this case we see that longer windows require less iterations to converge. Each iteration becomes more efficient, but this increase of efficiency is not sufficient to compensate the increase of the CPU time necessary to perform each iteration. So, even in 4D-VAR  the total CPU time   is  bigger for longer windows. 

When the assimilated data are noisy, more data results in a better accuracy of identified topography. It was shown that smaller time step and proportionally smaller assimilation window allows us to obtain the same  precision in the reconstructed topography. However, the dependence of the residual error in topography on the quantity of assimilated data is not uniform for all windows. 
When assimilation window is shorter than 2 days, the residual error decreases as inverse of the square root of the window length. But when we assimilate data with windows longer than 2 days, the decrease rate of the residual  becomes proportional to $T^{-3/2}$.

When the noise source  is considered, we see the most dangerous noise lies  in initial conditions. The same amplitude of noise in the forcing of the model and in its initial point may result in 30 times bigger error in topography in the second case. Consequently, one can think about data assimilation for the joint  simultaneous identification of  topography and  initial point. On the other hand, the final result exhibits relatively low sensitivity to measurements errors and noise in the assimilated data. 

We can state  the  topography identified once by data assimilation is valid for other model runs that start from other initial conditions and are situated in other parts of the model's attractor. The reconstructed topography can  be used in the model for sufficiently long runs that exceed many times the length of the assimilation window. The final model's error due to inexact topography reconstructed assimilating noisy data depends linearly on the noise amplitude in these data. 

Thus, we can state it is  possible  to use the  assimilation of external data in order to reconstruct the bottom topography of a nonlinear barotropic  model in the case of the time dependent motion.  One open question at this time concerns   possible principal difficulties related to baroclinicity and multi-layer models. In particular, optimization of topography may result in modification of the geometry of the basin at certain layers. Multi-layer model may also present theoretical particularity and invoke the question about differentiability of the  model with respect to the topographic field. 

Another point that has not been discussed here, is the possible lack of external data. We have supposed all the necessary data are available, sea surface elevation and all velocity fields. In practice, however, only the surface height can be easily measured and assimilated. 

And finally, it must by noted that the use of the bottom topography as control is only one example of a non traditional control variable. A number of model parameters,  may require such an optimization. One of them, an optimal  choice of boundary conditions on rigid and open boundaries seems  to be the first necessity.

Acknowledgments.
All the contour pictures have been prepared by the Grid Analysis and Display System (GrADS) developed in  the Center for Ocean-Land-Atmosphere Interactions,   Department of Meteorology, University of Maryland.

Author is grateful to two anonymous reviewers for truly valuable comments
 and constructive suggestions on the manuscript. Their insightful reviews have
 enabled him to improve upon the original manuscript.

\bibliography{/global/users/kazan/text/mybibl}
 
\mkpicstoend 

\end{document}